Sensor applications

# Sensor-Based Continuous Hand Gesture Recognition by Long Short-Term Memory

Tsung-Ming Tai, Yun-Jie Jhang, Zhen-Wei Liao, Kai-Chung Teng, and Wen-Jyi Hwang

[1]Department of Computer Science and Information Engineering, National Taiwan Normal University, Taipei 117, Taiwan

Abstract— This article aims to present a novel sensor-based continuous hand gesture recognition algorithm by long short-term memory (LSTM). Only the basic accelerators and/or gyroscopes are required by the algorithm. Given a sequence of input sensory data, a many-to-many LSTM scheme is adopted to produce an output path. A maximum *a posteriori* estimation is then carried out based on the observed path to obtain the final classification results. A prototype system based on smartphones has been implemented for the performance evaluation. Experimental results show that the proposed algorithm is an effective alternative for robust and accurate hand-gesture recognition.

## I. INTRODUCTION

A continuous hand gesture is a sequence of hand movements conducted by a human. Continuous hand gesture recognition techniques are important in human–machine interaction (HMI) and human activity recognition (HAR). They find applications in many areas, such as smart home, health care, augmented reality, virtual reality, machine control, and sign-language recognition [1]. A common approach for continuous hand gesture recognition is to adopt vision-based gestures recognition (VGR) techniques. A VGR method is usually based on captured video sequences by one or several cameras for interpreting and analyzing the motion [1]–[4]. Although some VGR techniques are effective, they may incur high computational load for online recognition, in which real-time operations on the video sequences are required.

An alternative to VGR techniques is a sensor-based gestures recognition (SGR) technique, which carries out continuous gesture recognition from the data produced by sensors. Accelerometers, gyroscopes, electromyography, and/or inertial sensors are the sensors used in many SGR-based systems [5]–[10]. Some of the sensors are commonly deployed in smart devices, such as smartphones and smart wear. With the proliferation of smart devices, the SGR techniques are emerging as the major approaches for HMI and HAR.

A challenging issue for SGR techniques is the gesture spotting in a sequence of gestures. The goal of gesture spotting is to identify start and end positions of gestures so that multiple continuous gestures can be recognized independently. In [7] and [9], a user action is required for gesture spotting. Dedicated sensors for spotting are employed in [8]. These approaches may introduce additional overhead for the gesture recognition. Continuous hand gesture recognition schemes offering automatic detection of start and end points of gestures are presented in [5], [6] and [10]. The detection is based on thresholding operations over the variances of sensory data. The performance of the SGR method may be sensitive to the selection of thresholds.

The long short-term memory (LSTM) [11] may be an alternative for the continuous hand gesture recognition. The LSTM is a deep learning [12] technique capable of exploiting both long-term and short-term behavior of input data for accurate classification. It is a variant of recurrent neural network and is based on a back propagation through time technique for training. The LSTM has been successfully applied to various applications, such as speech recognition, image caption generation, and machine translation [13]. In addition, a number of LSTM techniques have been proposed for continuous hand gesture recognition. The LSTM algorithms in [14] and [15] carry out VGR-based hand-gesture recognition in conjunction with a convolutional neural network. The LSTM algorithms can also operate along for solving SGR recognition problems [15]. Although the algorithms have been found to be effective, many techniques have a common drawback that they are designed only for a single continuous gesture. The gesture-spotting issue is still not addressed.

This aims to present a novel SGR technique based on LSTM. Only the basic accelerometers and gyroscopes commonly used in smartphones or devices are required for the hand-gesture recognition. This may facilitate the deployment of the algorithm to large varieties of the applications in smart devices. The technique is able to carry out the recognition of multiple continuous gestures from the sensory data by the LSTM. It is based on a many-to-many inference scheme, where an output is produced by the LSTM at each time step. A collection of output sequence is viewed as an output path. A simple gesture-spotting scheme is then carried out to find the maximum *a posteriori* (MAP) solution for the final classification, given the observed output path.

A prototype system of the proposed algorithm based on the smartphones has been developed for performance evaluation. Experimental results reveal that the proposed algorithm is well suited for the continuous hand gesture recognition applications on the smart devices requiring high robustness and high classification accuracy.

## II. PROPOSED ALGORITHM

The proposed algorithm is based on LSTM [12], which takes an input sequence $X = \{x_t, t = 1, ..., T\}$, and computes the hidden states $h_t$, memory cells $c_t$, and the output sequence $y_t$ from $t = 1$ to $t = T$ iteratively. At each time step $t$, it first computes $c_t$ and $h_t$ from $x_t$, $c_{t-1}$, and $h_{t-1}$, where the initial hidden state $h_0$ and initial memory cell $c_0$ are vectors of zeros. The output $y_t$ is then obtained from the hidden state $h_t$.

The LSTM can be directly used for the continuous hand gesture recognition. In this case, $x_t$ are the input sensory data, and $y_t$ are the recognition results at time step $t$. Given a gesture sequence $X = (x_1, ..., x_T)$, the LSTM is served as a sequence-to-sequence predictor. It produces an output sequence $Y = (y_1, ..., y_T)$ based on an input sequence $X = (x_1, ..., x_T)$. All the elements in the sequence $X$ are vectors. They have an identical dimension $N$, which is dependent on the sensors adopted for the hand-gesture recognition.

Each element of the output sequence $Y$ is also a vector. Let $Q$ be the total number of gestures. Each $y_t \in Y$ has dimension $Q$. The softmax operator for computing $y_t$ given $h_t$ [12] can be viewed as an estimation of probability $\Pr(a_t = j/h_t)$ for the occurrence of gesture $j$ given the hidden state $h_t$, where $a_t$ is the label produced by the proposed algorithm at time step $t$. Let $y_{t,j}$ be the $j$th element of $y_t$, $j = 1, ..., Q$. Therefore, we set $y_{t,j} = \Pr(a_t = j/h_t)$ in the proposed algorithm. Suppose $j_t$ is the index of the gesture having the largest probability at time step $t$. In other words

$$j_t = \arg\max_{1 \leq j \leq Q} y_{t,j} = \arg\max_{1 \leq j \leq Q} \Pr(a_t = j/h_t). \quad (1)$$

We then let $a_t = j_t$. Define the set $A = \{a_1, ..., a_T\}$ as the set of labels for the time steps $t = 1, ..., T$. We call $A$ a path given the input sequence $X$. In the proposed algorithm, the path is further mapped to a classification outcome $R = \{r_1, ..., r_k\}$ for the recognition of $k$ distinctive gestures, where $r_j$ is the $j$th gesture recognized by the algorithm. The mapping, denoted by $B(A)$, is based on the conditional probability model $\Pr(R/A)$ given by

$$\Pr(R/A) = \prod_{j=1}^{k} \Pr(r_j/A). \quad (2)$$

In the model, the probability $\Pr(r_j/A)$ can be evaluated as

$$\Pr(r_j/A) = \frac{|I_{r_j}|}{T} \quad (3)$$

where $I_i = \{t : a_t = i\}$, and $|I_i|$ indicates the cardinality of the set. We can view $I_i$ as the set of time steps where the gesture $i$ is the recognized gesture. The search for $I_i$ can be viewed as a gesture-spotting operation for the gesture $i$. Moreover, the mapping $B$ is an MAP estimator, which selects $R$ maximizing $\Pr(R/A)$ as the final classification outcome. That is

$$B(A) = \arg\max_{R} \Pr(R/A). \quad (4)$$

Let $\mathcal{K}$ be the set of gestures whose cardinality is among the top $k$ of the $Q$ gestures. From (2) and (3), we see that the $B(A)$ in (4) can be implemented by setting $r_j = i$, where the gesture $i$ is the $j$th gesture occurred in the path $A$ among the gestures in $\mathcal{K}$. Fig. 1 shows an example of mapping from path $A = \{a_t, 1 \leq t \leq T\}$ to final classification outcome $R = \{r_1, r_2, r_3\}$ for the recognition of three gestures (i.e., $k = 3$). In this example, $\mathcal{K}$ contains gestures 1, 2, and 3.

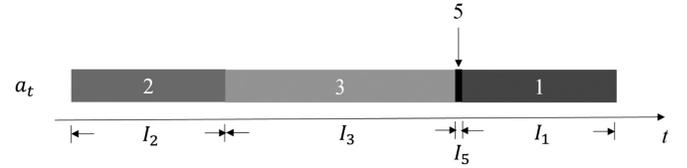

Fig. 1. Example of mapping from path $A = \{a_t, 1 \leq t \leq T\}$ to final classification outcome $R = \{r_1, r_2, r_3\}$ for the recognition of three gestures (i.e., $k = 3$). In this example, $r_1 = 2$, $r_2 = 3$, and $r_3 = 1$.

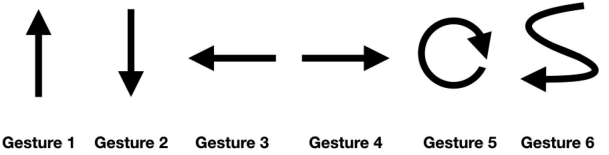

Gesture 1  Gesture 2  Gesture 3  Gesture 4  Gesture 5  Gesture 6

Fig. 2. Gestures considered in the experiments.

TABLE 1. Gestures and Their Actions for Various Home Appliances.

|  | Gest. 1 | Gest. 2 | Gest. 3 | Gest. 4 | Gest. 5 | Gest. 6 |
|---|---|---|---|---|---|---|
| TV | Volume Up | Volume Down | Previous Chan. | Next Chan. | Play or Pause | Video Source |
| Air Cond. | Temp. Up | Temp. Down | Air Vol. Up | Air Vol. Down | Play or Pause | Func. Select. |

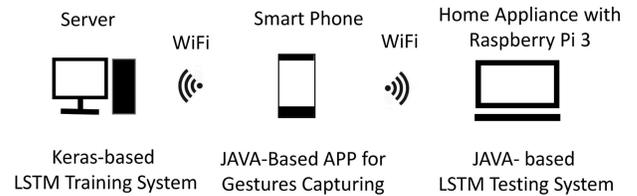

Server — Keras-based LSTM Training System  
Smart Phone — JAVA-Based APP for Gestures Capturing  
Home Appliance with Raspberry Pi 3 — JAVA-based LSTM Testing System

Fig. 3. Setup of the experiments.

The order of occurrence in path $A$ for the gestures in $\mathcal{K}$ is gestures 2, 3, and 1. Consequently, $r_1 = 2$, $r_2 = 3$, and $r_3 = 1$.

## III. EXPERIMENTAL RESULTS

There are six classes of gestures considered in the experiments, as shown in Fig. 2. The dimension of each output $y_t$ is then 6 (i.e., $Q = 6$). The gestures can be applied for the remote control of home appliances, as suggested in [5]. Examples of the gestures and their corresponding actions for some home appliances are included in Table 1.

Fig. 3 shows the setup of the experiments. As shown in the figure, a dedicated server is adopted for LSTM training, which is carried out offline by Keras [16] with backend Tensorflow. The server is a personal computer with Intel I7 CPU and Nvidia GTX 1070 GPU. A platform different from training server is adopted for LSTM testing. It is a simple Raspberry Pi 3 embedded system for the control of home appliances. The LSTM testing system is implemented by JAVA. It is built from the LSTM parameters acquired from Keras after training process are completed.

All the gesture sequences for training and testing are captured by a smartphone equipped with accelerometer and gyroscope. They are capable of measuring acceleration and angular velocity in three orthogonal axes, respectively. The dimension of each sample $x_t$ in each training or testing sequence is then $N = 6$. The smartphone considered in the experiments is the HTC One M9 operating in Android system. A JAVA-based APP is built for gesture capturing and delivery, where the number of gestures $k$ can be prespecified. The selection of $k$ is application oriented. We set $k = 1$ for a single action, and $k > 1$ when multiple actions are desired. Figs. 4 and 5 show the samples of waveforms produced by accelerator and gyroscope for each gesture, respectively. The results of measurement are delivered to the training server or home appliance by WiFi for LSTM training and testing operations, respectively.

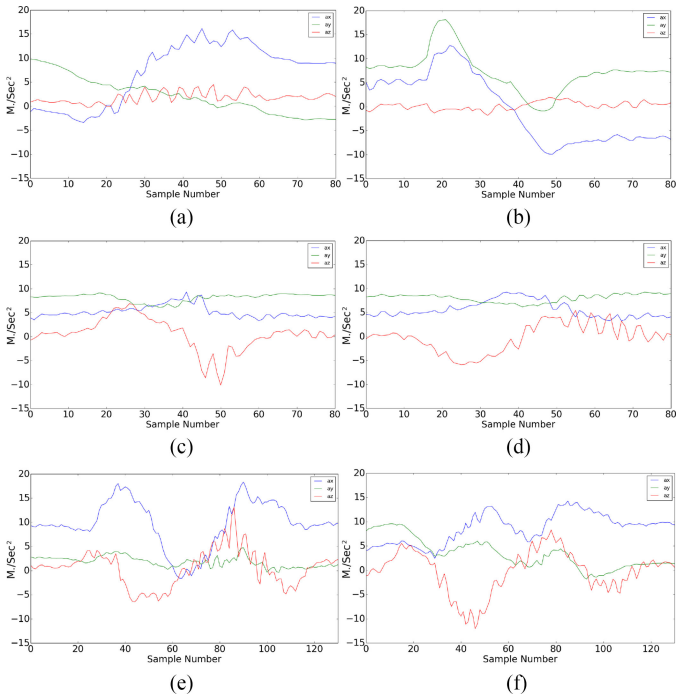

Fig. 4. Samples of waveforms produced by accelerator in three axes (ax, ay, and az) for each gesture. (a) Gesture 1. (b) Gesture 2. (c) Gesture 3. (d) Gesture 4. (e) Gesture 5. (f) Gesture 6.

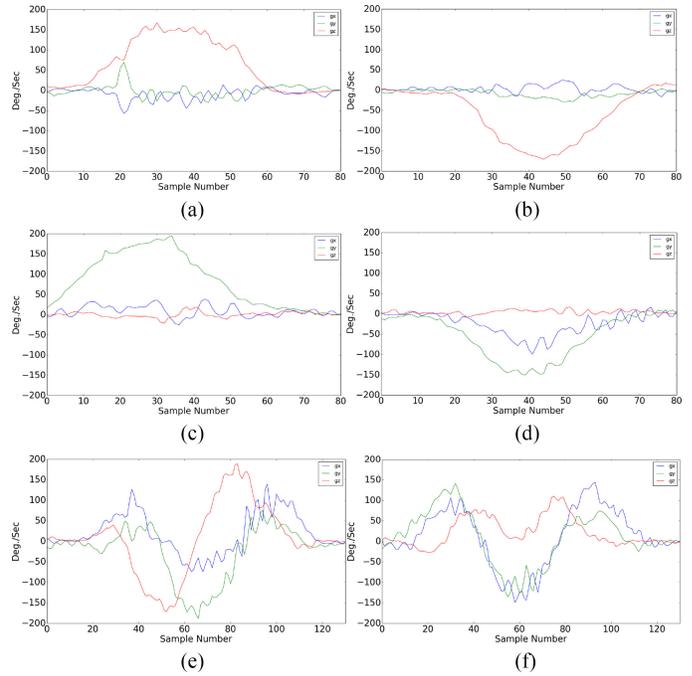

Fig. 5. Samples of waveforms produced by gyroscope in three axes (gx, gy, and gz) for each gesture. (a) Gesture 1. (b) Gesture 2. (c) Gesture 3. (d) Gesture 4. (e) Gesture 5. (f) Gesture 6.

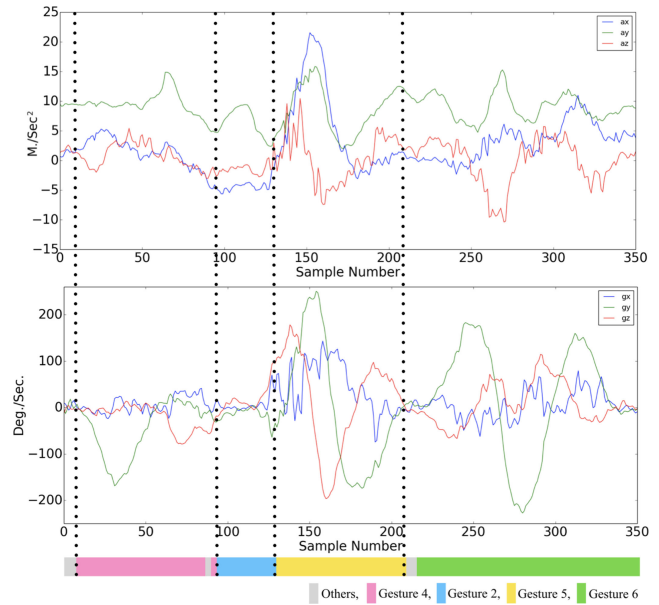

Fig. 6. Example of a test sequence produced by accelerator and gyroscope consisting of four hand gestures (Gesture 4, Gesture 2, Gesture 5, and Gesture 6) back-to-back. The gesture-spotting results are also shown in the bottom of the figure.

The training dataset contains 2900 sequences produced by five participants including experienced users and new users. The sequences acquired from the experienced users are collected at different days for capturing different variations in gesturing, such as variations in lengths of gestures and/or amplitudes of samples. The collection of sequences form each new user is completed in a single day. The variations due to the unfamiliarity to the gesture movement can then be recorded for training.

The evaluation of the proposed algorithm is based on the LSTM testing system in Fig. 3 with the testing dataset different from the training set. There are 3100 hand gestures in the testing set produced by seven participants. Each test sequence may contain two or more hand gestures. The initial orientation of smartphone for data acquisition of both training and testing sequences is portrait orientation. Moreover, some testing sequences contain additional vibration due to shaking or unstable hand. Fig. 6 shows an example of a testing sequence produced by accelerator and gyroscope consisting of four hand gestures (Gesture 4, Gesture 2, Gesture 5, and Gesture 6) back-to-back. The results of gesture spotting are also revealed in the bottom of the figure. It can be observed from Fig. 6 that direct gesture spotting is difficult, even by visual inspection. Nevertheless, we can see from the bottom of Fig. 6 that the sets $I_4$, $I_2$, $I_5$, and $I_6$ are the sets having the largest cardinality. Therefore, the recognition outcome is $r_1 = 4$, $r_2 = 2$, $r_3 = 5$, and $r_4 = 6$. Accurate recognition for the continuous gestures can be achieved.

For the LSTM, the dimension of memory cell $c_t$ and hidden state $h_t$ are 32**. Both $c_t$ and $h_t$ are responsible for abstracting the input sensory data for classification. The selection of different dimensions may result in a different classification accuracy, which is defined as

TABLE 2. Confusion Matrix on the Test Dataset for the Proposed Algorithm.

|        | Gest. 1 | Gest. 2 | Gest. 3 | Gest. 4 | Gest. 5 | Gest. 6 |
|--------|---------|---------|---------|---------|---------|---------|
| Gest. 1 | 96.95 | 1.66  | 0.00  | 0.00  | 1.39   | 0.00  |
| Gest. 2 | 1.37  | 95.22 | 2.56  | 0.00  | 0.68   | 0.17  |
| Gest. 3 | 0.36  | 0.00  | 99.28 | 0.00  | 0.00   | 0.36  |
| Gest. 4 | 0.28  | 0.00  | 0.85  | 97.17 | 0.00   | 1.70  |
| Gest. 5 | 0.00  | 0.00  | 0.00  | 0.00  | 100.0  | 0.00  |
| Gest. 6 | 0.21  | 0.00  | 0.00  | 0.00  | 0.21   | 99.59 |

Each cell in the confusion matrix represents the percentage in which the gesture in the corresponding row is classified as the gesture in the corresponding column.

TABLE 3. Hit Rate in Percentage of the Proposed Continuous Gesture-Recognition Algorithms with Only Accelerator or Gyroscope, or Both.

|            | $H_1$ | $H_2$ | $H_3$ | $H_4$ | $H_5$ | $H_6$ |
|------------|-------|-------|-------|-------|-------|-------|
| accelerator | 84.76 | 43.52 | 86.36 | 95.18 | 98.53  | 78.79 |
| gyroscope   | 90.30 | 98.12 | 77.78 | 93.77 | 100.0  | 79.79 |
| both        | 96.95 | 95.22 | 99.28 | 97.17 | 100.00 | 99.59 |

TABLE 4. Hit Rate in Percentage of Various SGR-Based Continuous Gesture-Recognition Algorithms.

|          | $H_1$ | $H_2$ | $H_3$ | $H_4$ | $H_5$  | $H_6$ |
|----------|-------|-------|-------|-------|--------|-------|
| Proposed | 96.95 | 95.22 | 99.28 | 97.17 | 100.00 | 99.59 |
| [5]      | 93.50 | 93.75 | 91.20 | 92.50 | 95.00  | 95.00 |
| [17]     | 88.00 | 82.00 | 86.00 | 77.00 | 57.00  | 43.00 |

the number of gestures correctly classified divided by the total number of gestures in the testing set. In our experiments, the accuracy is 95.85%, 98.04%, and 97.27% when the dimension is 16, 32, and 64, respectively. Therefore, dimension 32 has a better accuracy. This is because the dimension 16 may be too small for including sufficient information for classification. On the contrary, the dimension 64 may be too large; therefore, noises from input data may also be included in the abstraction.

Table 2 shows the confusion matrix on the testing dataset for the proposed algorithm. The confusion matrix contains information about actual and predicted gesture classifications carried out by the system. Each cell in the confusion matrix represents the percentage in which the gesture in the corresponding row is classified as the gesture in the corresponding column. Let $H_i$ be the hit rate of gesture class $i$, which is defined as the number of gestures in class $i$ that are correctly classified divided by the total number of gestures in class $i$. Therefore, $H_i$ of the proposed algorithm is identical to the value of the cell in the $i$th column and the $i$th row of the confusion matrix shown in Table 2. It can be observed from the table that the proposed algorithm attains high hit rate $H_i$ for each gesture class $i = 1, ..., 6$.

The hit rates of the proposed algorithm will be degraded when only accelerator or gyroscope is used. Table 3 reveals the degradation in hit rate $H_i$ of each gesture class $i$ of the proposed algorithm with the employment of only one type of sensor. It can be observed from Table 3 that the hit rates of most gesture classes are reduced without the employment of both sensors. In particular, $H_6$ is degraded from 99.59% to 78.79% and 79% when only accelerator or gyroscope is used, respectively. These results show that the employment of both accelerator and gyroscope is beneficial for hand-gesture recognition.

Table 4 compares the hit rate $H_i$ of various SGR techniques. Direct comparisons may be difficult because these SGR techniques are based on different data-acquisition systems and training sets. Nevertheless, because the target gestures to be recognized are identical for the techniques, they could be applied to the same applications, such as remote control of home appliances [5]. It can be observed from Table 4 that the proposed algorithm could have superior performance over the techniques in [5] and [17] for smart-home applications.

## IV. CONCLUDING REMARKS

A smartphone-based prototype has been built for performance evaluation. Experimental results reveal that the hit rate for all the gesture classes are above 95% for a test set consisting of 3100 gestures. As compared with the existing SGR-based gesture-recognition techniques, the proposed algorithm provides superior classification results. This is because the proposed algorithm is able to perform effective gesture spotting, even for continuous gestures back-to-back. It is therefore beneficial for HMI or HAR applications where reliable continuous hand gesture recognition is desired. A possible extension of the proposed work is to accommodate custom gestures for providing flexibilities for a user to meet the needs of a particular application.